\documentclass[12pt]{article}
\usepackage{amsmath,amsthm,amscd,array,amssymb}
\begin{document}
{\large

\begin{center}
{\Large{\bf Fedosov supermanifolds: II. Normal coordinates}}

\vspace{1cm}

{\sc Bodo~Geyer}$^{\ a)}\ $\footnote{E-mail: geyer@itp.uni-leipzig.de} and
{\sc Peter~Lavrov}$^{\ a)\ b)}\ $\footnote{E-mail: lavrov@tspu.edu.ru}

\vspace{.5cm}
{\normalsize\it $^{a)}$ Center of Theoretical Studies and Institute
of Theoretical Physics,\\ Leipzig University,
Augustusplatz 10/11, D-04109 Leipzig, Germany}

\vspace{.3cm}

{\normalsize\it $^{b)}$ Tomsk State Pedagogical University,
634041 Tomsk, Russia}

\end{center} }

\vspace{.5cm}

\begin{quotation}
\noindent
\normalsize
The study of recently introduced Fedosov supermanifolds \cite{gl2} is continued.
Using normal coordinates, properties of even and odd symplectic
supermanifolds endowed with a symmetric connection respecting given sympletic structure are studied.
\end{quotation}

\section{Introduction}
The formulation of fundamental physical theories, classical as well as quantum ones,
by differential geometric methods nowadays is well established and has a great conceptual virtue.
Probably, the most prominent example is the formulation of general relativity on Riemannian manifolds,
i.e., the geometrization of the gravitational force; no less important is the geometric
formulation of gauge field theories of the fundamental forces on fiber bundles. Another essential
route has been opened by the formulation of classical mechanics -- and also classical field theories -- on
symplectic manifolds and their connection with the (geometric) quantization. The properties of such kind of
manifolds are widely studied.

Recently, some specific considerations in quantum field theory involve more complicated manifolds namely
the so-called Fedosov manifolds, i.e., symplectic manifolds equipped with a symmetric connection which
respects the symplectic structure. Fedosov manifolds have been introduced for the first time in the framework
of deformation quantization \cite{F}. The properties of Fedosov manifolds have been investigated in details
(see, e.g.,~Ref.~\cite{fm}). Especially, let us mention that for any Fedosov manifold the scalar curvature $K$
is trivial, $K=0$, and that the intrigue relation $\omega_{ij,kl}=(1/3)R_{klij}$, in terms of normal coordinates,
holds between the symplectic structure $\omega_{ij}$ and the curvature tensor $R_{klij}$.

The discovery of supersymmetry \cite{GL} has introduced into
modern quantum field theory the notion of supermanifolds proposed
by Berezin \cite{Ber}. Systematic considerations of supermanifolds
and Riemannian supermanifolds were performed by DeWitt
\cite{DeWitt}. At present, even and odd symplectic supermanifolds
and the corresponding differential geometry are widely
involved and studied in consideration of some problems of modern theoretical and
mathematical physics \cite{Wit}.

However, the situation concerning Fedosov supermanifolds is quite
different. Only some specific problems concerning even Fedosov
supermanifolds were under consideration. Especially, flat even
Fedosov supermanifolds have been used in the study of a
coordinate-free scheme of deformation quantization \cite{bt}, for
an explicit realization of the extended antibrackets \cite{GrS}
and for the formulation of modified triplectic quantization in
general coordinates \cite{gl1}. Deformation quantization for any
even symplectic supermanifolds has been constructed in
Ref.~\cite{B} (see also \cite{BHW}). It was the main aim in
\cite{gl2} to start a systematic investigation of basic properties
of even and odd Fedosov supermanifolds. In particular, some basic
difference in even and odd Fedosov supermanifolds has been found
which can be expressed in terms of the scalar curvature $K$.
Namely, for any even Fedosov supermanifold the scalar curvature,
as in usual differential geometry, is equal to zero while for odd
Fedosov supermanifolds it is, in general, non-trivial. This
fundamental fact has forced us to continue the investigation of
the geometry of Fedosov supermanifolds. Thereby, without reviewing
it here, definitions and results of \cite{gl2} will be used where
we need them.

The paper is organized as follows. In Sec.~2, we consider the properties of
the Christoffel symbols in normal coordinates and derive an infinite system
of equations for the affine extensions of the Christoffel symbols. In Sec.~3,
the relation between the first order affine extension of Christoffel symbols and the symplectic curvature tensor both in normal and any local coordinates are studied. In Sec.~4, we derive a relation between the second order affine extension of the symplectic structure and the symplectic curvature tensor.
Sec.~5 is devoted to the derivation of identities containing the first order affine extension of the symplectic curvature tensor and to a relation between the third order affine extension of symplectic structure and the first order affine extension of the symplectic curvature tensor. In Sec.~6, a short summary is given.

We use the condensed notations and the definition of tensor fields on supermanifolds as given by DeWitt \cite{DeWitt}. Derivatives
with respect to the coordinates $x^i$ are understood as acting from the right and for them the notation $A_{,i}={\partial_r A}/{\partial x^i}$ is used. Covariant derivatives also act from the right $A_{;i}=A\nabla_i$.  The Grassmann
parity of any quantity $A$ is denoted by $\epsilon (A)$.


\section{Affine extensions of Christoffel symbols and tensors on symplectic supermanifolds}

Let us consider a $2N$-dimensional symplectic supermanifold $(M,\omega)$ which
is called even (odd) iff the non-degenerate closed 2-form $\omega$ is even (odd), i.e., $\epsilon(\omega)=0 \quad (\epsilon(\omega)=1)$. Let $\Gamma$ be a symmetric (affine) connection on $M$ (not necessarily preserving $\omega$).
Given a point $p \in M$ and, in a vicinity of $p$, local coordinates $\{x^i, \,
\epsilon (x^i)=\epsilon_i, i=1,...,2N \}$ then a geodesic $x(t)$ through $p$ is defined by
\begin{eqnarray}
\label{geodesic}
&&\frac{d^2 x^i}{dt^2} + \Gamma^i_{\;jk}
\frac{d x^k}{dt}\frac{d x^j}{dt}=0,\quad x(0)=p,
\quad
\epsilon(\Gamma^i_{\;jk})=\epsilon_i+\epsilon_j+\epsilon_k,\\
\nonumber
&&\Gamma^i_{\;jk}=(-1)^{\epsilon_j\epsilon_k}\Gamma^i_{\;kj},
\end{eqnarray}
with $t$ being a (real) canonical parameter along that geodesic.

The coordinate system is called {\em geodesic} (at $p$) iff
$\Gamma^i_{\;jk}(p) = 0$.
This means that for any geodesic coordinates $\{x^{\prime j}\}$ it holds
$x^i =  {\delta^i}_j x^{\prime j} + \phi^i(x^{\prime})$ where
$\phi^i(x^{\prime})$ together with its first two derivatives vanishes.
Let $T_p\,M$ be the tangent space at $p \in M$. Now, let
$v^i := \left(\frac{d x^i}{dt}\right)_{t=0} \in T_p\,M$ be the tangent
on the geodesic at $p$ which, together with $x(0) = p$
uniquely determines the geodesic curve. Then, in the vicinity of $p$, the
coordinate system $\{y^{i} = v^i\,t\}$ constitutes a particular class of geodesic coordinates being called {\em normal}.

In normal coordinates the geodesic equation (\ref{geodesic}) reduces,
\begin{eqnarray}
\label{Chsy}
 \Gamma^i_{\;jk}(y)\,y^k\,y^j = 0,
\end{eqnarray}
where $\Gamma^i_{\;jk}(y)$ are the Christoffel symbols being calculated
at the point $p$ in terms of the normal coordinates $y^i$. Conversely,
for any coordinate system $\{y^{i}\}$ obeying (\ref{Chsy}) any geodesic, due to
Eq.~(\ref{geodesic}), satisfies ${d^2 y^i}/{dt^2}= 0$ and, hence,
$\{y^{i} = v^i\,t\}$ with some
$v^i$.\footnote{
Here, we generalized to the case of supermanifolds the introduction of normal coordinates as has been given for affine manifolds for the first time by Veblen, see, e.g., \cite{E}. However, also the more modern definition \cite{W} using the exponential map $\exp_p \,: U \rightarrow M $ from a small neighborhood $U$ of $0 \in T_p\,M$ onto $M$ with the property $\exp_p(v) = x(1)$ could have been generalized. Thereby, the exponential map on the supermanifold $M$ has to be defined according to, e.g., DeWitt \cite{DeWitt}.
}
Also in case of supermanifolds, like in ordinary manifolds, it holds that under an arbitrary analytic transformation of the coordinates $x^i \rightarrow x^{\prime i}$ the corresponding normal coordinates undergo a linear homogeneous transformation with constant coefficients ${a^i}_j$,
\begin{eqnarray}
y^{\prime i } = {a^i}_j\ y^j.
\end{eqnarray}
In normal coordinates many of the geometric properties of the (super)manifold are much easier to derive then in arbitrary coordinates.\footnote{
Remind that in normal coordinates the equations of geodesics through the origin (at $p$) have the same form as the equations of straight lines in Euclidean space in cartesian coordinates.
}
This is the reason for using them in the following.

Obviously, normal coordinates are defined by the connection (and do not depend on $\omega$). However, the defining identities (\ref{Chsy}) can be rewritten in the equivalent form with the help of $\omega$,
\begin{eqnarray}
\label{Chsy1}
\Gamma_{ijk}(y)\,y^k\,y^j\equiv 0,
\end{eqnarray}
where, independent of the chosen coordinate system,
\begin{eqnarray}
\label{Chsyd}
&&\Gamma_{ijk}=\omega_{il}\Gamma^l_{\;\;jk},\qquad
\Gamma^i_{\;\;jk}=\omega^{il}\Gamma_{ljk}
(-1)^{\epsilon_l+\epsilon(\omega)(\epsilon_i+\epsilon_l)}, \\
\nonumber
&&\epsilon(\Gamma_{ijk})=\epsilon(\omega)+\epsilon_i
+\epsilon_j+\epsilon_k.
\end{eqnarray}
and
\begin{eqnarray}
\label{omega}
&&\omega^{ik}\;\omega_{kj}\,
(-1)^{\epsilon_k+\epsilon(\omega)(\epsilon_i+\epsilon_k)}
=\delta^i_j,\qquad
\omega_{ik}\;\omega^{kj}\,
(-1)^{\epsilon_i+\epsilon(\omega)(\epsilon_i+\epsilon_k)}
=\delta^j_i,
\end{eqnarray}
It follows from (\ref{Chsy1}) and the symmetry properties of $\Gamma_{ijk}$
w.r.t.~$(j\,k)$ that
\begin{eqnarray}
\label{Chsy0}
\Gamma_{ijk}(0)= 0.
\end{eqnarray}

In normal coordinates there exist additional relations at $p$ containing the partial derivatives of $\Gamma_{ijk}$. Namely, consider the Taylor expansion of $\Gamma_{ijk}(y)$ at $y=0$,
\begin{eqnarray}
\label{TCh}
\Gamma_{ijk}(y)=\sum_{n=1}^{\infty}\frac{1}{n!}
A_{ijkj_1...j_n}y^{j_n}\cdot\cdot\cdot y^{j_1},
\end{eqnarray}
where
\begin{eqnarray}
\label{Nt}
A_{ijkj_1\ldots j_n}=A_{ijkj_1\ldots j_n}(p)=\left.
\frac{\partial_r^n \Gamma_{ijk}}
{\partial y^{j_1}\ldots \partial y^{j_n}}\right|_{y=0}
\end{eqnarray}
is called an {\em affine extension} of $\Gamma_{ijk}$ of order $n=1,2,\ldots$ .
The symmetry properties of $A_{ijkj_1\ldots j_n}$ are evident from their definition (\ref{Nt}), namely, they are (generalized) symmetric w.r.t.~$(j\,k)$ as well as $(j_1\ldots j_n)$.

The set of all affine
extensions of $\Gamma_{ijk}$ uniquely defines a symmetric connection according to (\ref{TCh}). In terms of $A_{ijkj_1\ldots j_n}$ the (defining) property (\ref{Chsy1}) can be represented by the following sequence of identities:
\begin{eqnarray}
\label{rn}
\nonumber
\hspace{-1cm}
&&A_{ijli_1\cdot\cdot\cdot i_n}
+\sum_{k=1}^{n}A_{iji_kli_1
\cdot\cdot\cdot i_{k-1}i_{k+1}\cdot\cdot\cdot i_n}
(-1)^{\epsilon_{i_k}(\epsilon_l+\epsilon_{i_1}+\cdot\cdot\cdot
+\epsilon_{i_{k-1}})}
\\
\hspace{-1cm}
&&+\sum_{k=1}^{n}
A_{ili_kji_1\cdot\cdot\cdot i_{k-1}i_{k+1}\cdot\cdot\cdot i_n}
(-1)^{\epsilon_j(\epsilon_l+\epsilon_{i_k})+
\epsilon_{i_k}(\epsilon_l+\epsilon_{i_1}+\cdot\cdot\cdot
+\epsilon_{i_{k-1}})}
\\
\nonumber
\hspace{-1cm}
&&+\sum_{k=1}^{n-1}\sum_{m=k+1}^{n}
A_{ii_ki_mjli_1\cdot\cdot\cdot i_{k-1}i_{k+1}\cdot\cdot\cdot
i_{m-1}i_{m+1}\cdot\cdot\cdot i_n}
(-1)^{(\epsilon_{i_k}+\epsilon_{i_m})(\epsilon_j+\epsilon_l+
\epsilon_{i_1}+\cdot\cdot\cdot +
\epsilon_{i_{k-1}})+\epsilon_{i_m}(\epsilon_{i_{k+1}}+
\cdot\cdot\cdot +\epsilon_{i_{m+1}})}=0.
\end{eqnarray}
For given $n$ these identities contain $P=1+2n+n(n-1)/2=(n+2)(n+1)/2$ terms.
For example, when $n=1$ we have $P=3$ terms:
\begin{eqnarray}
\label{r3}
A_{ijkl}+A_{ijlk}(-1)^{\epsilon_k\epsilon_l} +
A_{iklj}(-1)^{\epsilon_j(\epsilon_l+\epsilon_k)}=0,
\end{eqnarray}
and for $n=2$ we have $ P=6$ terms:
\begin{eqnarray}
\label{r6}
\nonumber
&&A_{ijklm}+
A_{ijlkm}(-1)^{\epsilon_k\epsilon_l} +
A_{ikljm}(-1)^{\epsilon_j(\epsilon_l+\epsilon_k)}\\
&&+A_{ijmkl}(-1)^{\epsilon_m(\epsilon_l+\epsilon_k)}+
A_{ilmjk}(-1)^{(\epsilon_j+\epsilon_k)(\epsilon_m+\epsilon_l)} +
A_{ikmjl}(-1)^{\epsilon_j(\epsilon_m+\epsilon_k)+\epsilon_m\epsilon_l}=0.
\end{eqnarray}

Analogously, the affine extensions of an arbitrary tensor
$T=(T^{i_1...i_k}_{\;\;\;\;\;\;\;\;\;\;m_1...m_l})$ on $M$ are defined as  tensors on $M$ whose components at $p\in M$ in the local
coordinates $(x^1,\ldots,x^{2N})$ are given by the formula
\begin{eqnarray}
\label{AfT}
T^{i_1...i_k}_{\;\;\;\;\;\;\;\;m_1...m_l,j_1...j_n} \equiv
T^{i_1...i_k}_{\;\;\;\;\;\;\;\;m_1...m_l,j_1...j_n}(0)=\left.
\frac{\partial_r^n
T^{i_1...i_k}_{\;\;\;\;\;\;\;\;\;m_1...m_l}}
{\partial y^{j_1}...\partial y^{j_n}}\right|_{y=0}
\end{eqnarray}
where $(y^1,\ldots,y^{2N})$ are normal coordinates associated with
$(x^1,\ldots,x^{2N})$ at $p$. The first extension of any tensor
coincides with its covariant derivative because
$\Gamma^i_{\;\;jk}(0)=0$ in normal coordinates.

In the following, any relation containing affine extensions are to be
understood as holding in a neighborhood $U$ of an arbitrary point $p \in M$.
Let us also observe the convention that, if a relations holds for arbitrary
local coordinates, the arguments of the related quantities will be suppressed.

\section{First order affine extension of Christoffel symbols
and curvature tensor of Fedosov supermanifolds}

Suppose we are given an even (odd) Fedosov supermanifold
$(M,\omega, \Gamma)$, $\epsilon(\omega)=0, \; (\epsilon(\omega)=1)$.
This means that the symmetric connection $\Gamma$ (or, equivalently, the covariant derivative $\nabla$) respects the symplectic structure $\omega:\; \omega\nabla=0$.
In local coordinates $(x)$ this condition reads
\begin{eqnarray}
\label{sc}
\omega_{ij,k}=\Gamma_{ijk}-\Gamma_{jik}(-1)^{\epsilon_i\epsilon_j}.
\end{eqnarray}
The curvature tensor $R_{ijkl}(x)$, $\epsilon(R_{ijkl})
= \epsilon(\omega)+\epsilon_i+\epsilon_j+\epsilon_k+\epsilon_l$
of an even (odd) symplectic connection $\Gamma$
has the following representation (see Ref.~\cite{gl2}),
\begin{align}
\label{R}
\nonumber
R_{ijkl}=&-\omega_{in}\Gamma^n_{\;\;jk,l}+\omega_{in}\Gamma^n_{\;\;jl,k}
(-1)^{\epsilon_k\epsilon_l}+
\Gamma_{ikn}\Gamma^n_{\;\;jl}(-1)^{\epsilon_k\epsilon_j}-
\Gamma_{iln}\Gamma^n_{\;\;jk}(-1)^{\epsilon_l(\epsilon_j+\epsilon_k)}\\
=&-\Gamma_{ijk,l}+\Gamma_{ijl,k}(-1)^{\epsilon_k\epsilon_l}+
\Gamma_{nik}\Gamma^n_{\;\;\;jl}
(-1)^{\epsilon_n\epsilon_i+\epsilon_k(\epsilon_j+\epsilon_n)}-
\Gamma_{nil}\Gamma^n_{\;\;\;jk}
(-1)^{\epsilon_n\epsilon_i+\epsilon_l(\epsilon_k+\epsilon_j+\epsilon_n)},
\end{align}
and obeys the following symmetry properties
\begin{eqnarray}
\label{Rans}
R_{ijkl}=-(-1)^{\epsilon_k\epsilon_l}R_{ijlk},\qquad
R_{ijkl}=(-1)^{\epsilon_i\epsilon_j}R_{jikl};
\end{eqnarray}
in deriving (\ref{R})
the following relations, due to (\ref{Chsyd}) and (\ref{omega}),
\begin{eqnarray}
\label{ur}
\Gamma^n_{\;\;jk,l}
&=&
(-1)^{\epsilon_m+\epsilon(\omega)(\epsilon_m+\epsilon_n)}
\left(\omega^{nm}\Gamma_{mjk,l}+\omega^{nm}_{\;\;\;\;\;\;,l}\Gamma_{mjk}
(-1)^{\epsilon_l(\epsilon(\omega)+\epsilon_k+\epsilon_j+\epsilon_m)}\right),
\\
\omega_{in}\,\omega^{nm}_{\;\;\;\;\;\;,l}(-)^{\epsilon(\omega)\epsilon_n}
&=&
\omega_{in,l}\,\omega^{nm}\,
(-1)^{\epsilon_l(\epsilon(\omega)+\epsilon_n+\epsilon_m)
+\epsilon(\omega)\epsilon_n}.
\end{eqnarray}
were used together with the Eq.~(\ref{sc}).

As has been shown in Ref.~\cite{gl2}, for the symplectic curvature tensor $R_{ijkl}$ there holds the (super) Jacobi identity,
\begin{eqnarray}
\label{Rjac1}
 R_{ijkl}(-1)^{\epsilon_j\epsilon_l}
+R_{iljk}(-1)^{\epsilon_l\epsilon_k}
+R_{iklj}(-1)^{\epsilon_k\epsilon_j}=0,
\end{eqnarray}
and the first Bianchi identity containing a cyclic permutation of all the indices,
\begin{eqnarray}
\label{1BI}
R_{ijkl}(-1)^{\epsilon_i\epsilon_l}
+R_{lijk}(-1)^{\epsilon_l\epsilon_k+\epsilon_l\epsilon_j}
+R_{klij}
(-1)^{\epsilon_k\epsilon_j+\epsilon_l\epsilon_j+\epsilon_i\epsilon_k}+
R_{jkli}(-1)^{\epsilon_i\epsilon_j+\epsilon_i\epsilon_k}=0.
\end{eqnarray}

It follows from (\ref{sc}) that among the affine extensions of
$\omega_{ij}$ and $\Gamma_{ijk}$ there must exist some relations.
To this end let us consider the Taylor expansion of $\omega_{ij}$
in the normal coordinates $(y^1,\ldots,y^{2N})$ at $p\in M$,
\begin{eqnarray}
\label{Tew}
\omega_{ij}(y)
=\sum_{n=1}^{\infty}\frac{1}{n!} \Omega_{ij,j_1...j_n}\,
y^{j_n}\cdots y^{j_1},\quad
\Omega_{ij,j_1...j_n}=\omega_{ij,j_1...j_n}(0).
\end{eqnarray}
Using the symmetry properties of $\omega_{ij,j_1 \ldots j_n}(0)$ one can
easily obtain the following Taylor expansion for $\omega_{ij,k}$
\begin{eqnarray}
\label{Tedw}
\omega_{ij,k}(y)
=\sum_{n=1}^{\infty}\frac{1}{n!}
\Omega_{ij,kj_1...j_n}\, y^{j_n}\cdots y^{j_1}.
\end{eqnarray}
Taking into account (\ref{sc}) and comparing
(\ref{TCh}) and (\ref{Tedw}) we obtain
\begin{eqnarray}
\label{wA}
\Omega_{ij,kj_1\ldots j_n}=A_{ijkj_1\ldots j_n}-
A_{jikj_1\ldots j_n}(-1)^{\epsilon_i\epsilon_j}.
\end{eqnarray}
In particular,
\begin{eqnarray}
\label{wA1}
\Omega_{ij,kl}=A_{ijkl}-
A_{jikl}(-1)^{\epsilon_i\epsilon_j}.
\end{eqnarray}

Now, consider the curvature tensor $R_{ijkl}$, Eq.~(\ref{R}),
in the normal coordinates at $p\in M$. Then, due to $\Gamma_{ijk}(p)=0$, we obtain the following representation of the curvature tensor in terms of the affine extensions of the symplectic connection
\begin{eqnarray}
\label{Rnc}
R_{ijkl}(0)= -
A_{ijkl}+A_{ijlk}(-1)^{\epsilon_k\epsilon_l}.
\end{eqnarray}
Taking into account (\ref{r3}) 
and (\ref{Rnc}) the existence of some relation containing the curvature
tensor and the second affine extension of $\omega$ can be expected.
Indeed, 
step by step one obtains
\begin{eqnarray}
\label{RA}
\nonumber
R_{iklj}(0)&=&A_{ikjl}(-1)^{\epsilon_l\epsilon_j}-A_{iklj}=
A_{ijkl}(-1)^{(\epsilon_l+\epsilon_k)\epsilon_j}-A_{iklj}\\
\nonumber
&=&A_{ijkl}(-1)^{(\epsilon_l+\epsilon_k)\epsilon_j}+
A_{ikjl}(-1)^{\epsilon_l\epsilon_j}+
A_{iljk}(-1)^{(\epsilon_l+\epsilon_j)\epsilon_k}\\
\nonumber
&=&2A_{ijkl}(-1)^{(\epsilon_l+\epsilon_k)\epsilon_j}+
A_{iljk}(-1)^{(\epsilon_l+\epsilon_j)\epsilon_k}\\
\nonumber
&=& 2A_{ijkl}(-1)^{(\epsilon_l+\epsilon_k)\epsilon_j}+
A_{ijlk}(-1)^{(\epsilon_l+\epsilon_j)\epsilon_k+\epsilon_j\epsilon_l}\\
\nonumber
&=&
2A_{ijkl}(-1)^{(\epsilon_l+\epsilon_k)\epsilon_j}+
[R_{ijkl}(0)+A_{ijkl}](-1)^{(\epsilon_l+\epsilon_k)\epsilon_j}\\
&=&
3A_{ijkl}(-1)^{(\epsilon_l+\epsilon_k)\epsilon_j}+
R_{ijkl}(0)(-1)^{(\epsilon_l+\epsilon_k)\epsilon_j}.
\end{eqnarray}
Therefore we have the representation of the first order affine
extension of
the Christoffel symbols in terms of the curvature tensor at $p$,
\begin{eqnarray}
\label{RA1}
A_{ijkl}\equiv \Gamma_{ijk,l}(0) =
-\frac{1}{3}\left[R_{ijkl}(0)+R_{ikjl}(0)(-1)^{\epsilon_k\epsilon_j}\right],
\end{eqnarray}
where the symmetry properties (\ref{Rans}) of the curvature tensor
were used.

Notice, that relation (\ref{RA1}) was derived in normal coordinates.
It seems to be of general interest to find its analog relation
in terms arbitrary local coordinates $(x)$ because the Christoffel symbols
are not tensors while the r.h.s. of (\ref{RA1}) is a tensor.
To this end let us relate the arbitrary local coordinates to the normal ones as follows \cite{E},
\begin{eqnarray}
\label{alc}
x^i = x_0^i + y^i - \frac{1}{2} \left({\Gamma^i}_{jk}\right)_0 y^k\,y^j
+ \cdots + \frac{1}{n!}
\left(\frac{\partial^n_r x^i}{\partial y^{j_1}\cdots \partial y^{j_n}}\right)_0
y^{j_n}\cdots y^{j_1} + \cdots \,,
\end{eqnarray}
where the subscript $0$ indicates that the corresponding quantity is taken at the point $p\in M$. As a consequence we have
\begin{eqnarray}
\label{part1}
\left(\frac{\partial_r x^i}{\partial y^j}\right)_0 = \delta^i_j.
\end{eqnarray}
Since the Jacobian at $p$ is different of zero the series (\ref{alc})
can be inverted,
$y^i = x^i - x_0^i + \phi^i(x - x_0)$,
where $\phi^i(x - x_0)$ depends on the second and higher powers of the differences $x^j - x_0^j$.

Under that change of coordinates
$(x)\rightarrow (y)$ in some vicinity $U$ of $p$ the Christoffel symbols $\Gamma^i_{\;\;jk}$ transform according to the rule
\begin{eqnarray}
\label{Chtr}
\Gamma^i_{\;\;jk}(y)=
\frac{\partial_r y^i}{\partial{x}^l}
\left(\Gamma^l_{\;\;mn}({x})
\frac{\partial_r {x}^n}{\partial y^k}
\frac{\partial_r {x}^m}{\partial y^j}
(-1)^{\epsilon_k(\epsilon_j+\epsilon_m)}
+
\frac{\partial^2_r {x}^l}{\partial y^j\partial y^k}\right).
\end{eqnarray}
These relations can be rewritten in the form
\begin{eqnarray}
\label{Chtr0}
\Gamma_{ijk}(y)=
\left(\Gamma_{pqr}({x})
\frac{\partial_r {x}^r}{\partial y^k}
\frac{\partial_r {x}^q}{\partial y^j}
(-1)^{\epsilon_k(\epsilon_j+\epsilon_q)}
+ \;\omega_{pq}({x})
\frac{\partial^2_r {x}^q}{\partial y^j\partial y^k}\right)
\frac{\partial_r {x}^p}{\partial y^i}
(-1)^{(\epsilon_k+\epsilon_j)(\epsilon_i+\epsilon_p)},
\end{eqnarray}
where we used the definition (\ref{Chsyd}) of $\Gamma_{ijk}$ and the transformation rule of the symplectic structure,
\begin{eqnarray}
\omega_{ij}(y) = \omega_{pq}(x)
\frac{\partial_r {x}^q}{\partial y^j}
\frac{\partial_r {x}^p}{\partial y^i}
(-1)^{\epsilon_j(\epsilon_p + \epsilon_i)}
\end{eqnarray}
>From (\ref{Chtr}) one can express the matrix of second derivatives
in the form
\begin{eqnarray}
\label{Chtr1}
\frac{\partial^2_r {x}^q}{\partial y^j\partial y^k}=
\frac{\partial_r {x}^q}{\partial y^l}\Gamma^l_{\;\;jk}(y)-
\Gamma^q_{\;\;lm}({x})\frac{\partial_r {x}^m}{\partial y^k}
\frac{\partial_r {x}^l}{\partial y^j}
(-1)^{\epsilon_k(\epsilon_j+\epsilon_l)}.
\end{eqnarray}
In particular at $p\in M$ ($y=0$), observing the equality
(\ref{part1}) we have the relation
\begin{eqnarray}
\label{Chtr2}
\left(\frac{\partial^2_r x^q}{\partial y^j\partial y^k}\right)_0
=
- \Gamma^q_{\;\;lm}(x_0)
\left(\frac{\partial_r x^m}{\partial y^k}\right)_0
\left(\frac{\partial_r x^l}{\partial y^j}\right)_0
(-1)^{\epsilon_k(\epsilon_j+\epsilon_l)}
\equiv - \Gamma^p_{\;\;jk}(x_0).
\end{eqnarray}

Differentiating (\ref{Chtr0}) with respect to $y$ we find
\begin{align}
\label{Chtr4}
\nonumber
\Gamma_{ijk,l}(y)
=&
\;\Gamma_{pqr;s}(x)
\frac{\partial_r x^s}{\partial y^l}
\frac{\partial_r x^r}{\partial y^k}
\frac{\partial_r x^q}{\partial y^j}
\frac{\partial_r x^p}{\partial y^i}
(-1)^{(\epsilon_j+\epsilon_k+\epsilon_l)(\epsilon_i+\epsilon_p)+
(\epsilon_k+\epsilon_l)(\epsilon_j+\epsilon_q)
+\epsilon_l(\epsilon_k+\epsilon_r)}\\
\nonumber
&+ \omega_{pq,r}(x)
\frac{\partial_r x^r}{\partial y^l}
\frac{\partial^2_r x^q}{\partial y^j\partial y^k}
\frac{\partial_r x^p}{\partial y^i}
(-1)^{(\epsilon_k+\epsilon_j)(\epsilon_i+\epsilon_p)+
\epsilon_l(\epsilon_i+\epsilon_j+\epsilon_k+\epsilon_q+\epsilon_p)}\\
\nonumber
&+ \omega_{pq}(x)
\frac{\partial^2_r x^q}{\partial y^j\partial y^k}
\frac{\partial^2_r x^p}{\partial y^i\partial y^l}
(-1)^{(\epsilon_k+\epsilon_j)(\epsilon_i+\epsilon_p)}+
\omega_{pq}(x)
\frac{\partial^3_r x^q}{\partial y^j\partial y^k\partial y^l}
\frac{\partial_r x^p}{\partial y^i}
(-1)^{(\epsilon_j+\epsilon_k+\epsilon_l)(\epsilon_i+\epsilon_p)},
\end{align}
where the covariant derivative (for arbitrary local coordinates) is defined by
\begin{eqnarray}
\label{Chc}
\Gamma_{pqr;s}
=
\Gamma_{pqr,s}- \Gamma_{pqn}\Gamma^n_{\;\;\;rs}
- \Gamma_{pnr}\Gamma^n_{\;\;\;qs}
  (-1)^{\epsilon_r(\epsilon_n+\epsilon_q)}
- \Gamma_{nqr}\Gamma^n_{\;\;\;ps}
  (-1)^{(\epsilon_r+\epsilon_q)(\epsilon_n+\epsilon_p)}.
\end{eqnarray}
Now, making use of Eqs.~(\ref{sc}) and (\ref{Chsyd}), and restricting to the point $p \in M$, i.e., taking  $y=0$, we get
\begin{align}
\Gamma_{ijk,l}(0)
=
\left(\Gamma_{ijk;l}(x_0)-\Gamma_{iln}(x_0)\Gamma^n_{\;\;\;jk}(x_0)
(-1)^{\epsilon_l(\epsilon_j+\epsilon_k)}\right)+
\omega_{iq}(x_0)
\left(\frac{\partial^3_r x^q}{\partial y^j\partial y^k\partial y^l}\right)_0.
\end{align}
Due to (\ref{Chtr4}) and the identity (\ref{r3}),
\begin{eqnarray}
\nonumber
\Gamma_{ijk,l}(0)+\Gamma_{ijl,k}(0)(-1)^{\epsilon_k\epsilon_l}+
\Gamma_{ikl,j}(0)(-1)^{\epsilon_j(\epsilon_l+\epsilon_k)}=0,
\end{eqnarray}
the matrix of third derivatives at $p$ obeys the following relation,
\begin{align}
\label{3w1}
\nonumber
\omega_{iq}(x_0)
\left(\frac{\partial^3_r x^q}{\partial y^j\partial y^k\partial y^l}\right)_0
&=
\frac{1}{3}\Big[
-  \Gamma_{ijk;l}
- \Gamma_{ijl;k}(-1)^{\epsilon_k\epsilon_l}
- \Gamma_{ikl;j}(-1)^{\epsilon_j(\epsilon_k+\epsilon_l)}
\\
&\qquad
+ \Gamma_{iln}\Gamma^n_{\;\;\;jk}(-1)^{(\epsilon_j+\epsilon_k)\epsilon_l}
+ \Gamma_{ijn}\Gamma^n_{\;\;\;kl}
+ \Gamma_{ikn}\Gamma^n_{\;\;\;jl}(-1)^{\epsilon_k\epsilon_j}
\Big](x_0).
\end{align}
With the help of (\ref{3w1}) we get the following transformation law for $\Gamma_{ijk;l}$
under change of coordinates at the point $p$
\begin{align}
\label{Chtr5}
\Gamma_{ijk,l}(0)=
 \Big[\Gamma_{ijk,l}(x_0)-\frac{1}{3}Z_{ijkl}(x_0)\Big],
\end{align}
with the abbreviation
\begin{eqnarray}
\label{Zx}
Z_{ijkl}&=&
\Gamma_{ijk;l}+\Gamma_{ijl;k}(-1)^{\epsilon_k\epsilon_l}+
\Gamma_{ikl;j}(-1)^{(\epsilon_k+\epsilon_l)\epsilon_j}\\
\nonumber
&&
+2\Gamma_{iln}\Gamma^n_{\;\;\;jk}
(-1)^{(\epsilon_k+\epsilon_j)\epsilon_l}-
\Gamma_{ikn}\Gamma^n_{\;\;\;jl}(-1)^{\epsilon_j\epsilon_k}-
\Gamma_{ijn}\Gamma^n_{\;\;\;kl}
\end{eqnarray}

In straightforward manner one can check that the relations (\ref{Chtr4})
reproduce the right transformation law for the curvature tensor,
\begin{align}
\label{Rtr}
\nonumber
R_{ijkl}(0)
=&-\Gamma_{ijk,l}(0)+\Gamma_{ijl,k}(0)
(-1)^{\epsilon_k\epsilon_l}\\
\nonumber
=&\;\Big[-\Gamma_{ijk;l}+\Gamma_{iln}\Gamma^n_{\;\;\;jk}
(-1)^{\epsilon_l(\epsilon_j+\epsilon_k)}
+\Gamma_{ijl;k}(-1)^{\epsilon_k\epsilon_l}
-\Gamma_{ikn}\Gamma^n_{\;\;\;jl}
(-1)^{\epsilon_j\epsilon_k}\Big](x_0)
\\
\nonumber
=&
\left(R_{pqrs}(x)
\frac{\partial_r x^s}{\partial y^l}
\frac{\partial_r x^r}{\partial y^k}
\frac{\partial_r x^q}{\partial y^j}
\frac{\partial_r x^p}{\partial y^i}
(-1)^{(\epsilon_j+\epsilon_k+\epsilon_l)(\epsilon_i+\epsilon_p)+
(\epsilon_k+\epsilon_l)(\epsilon_j+\epsilon_q)
+\epsilon_l(\epsilon_k+\epsilon_r)}\right)_{x=x_0}
\\
=& R_{ijkl}(x_0),
\end{align}
where we have taken into account the following relation
\begin{align}
\nonumber
-\Gamma_{pqs;r}&+\Gamma_{prn}\Gamma^n_{\;\;\;qs}
(-1)^{\epsilon_r(\epsilon_q+\epsilon_s)}
+\Gamma_{pqr;s}
(-1)^{\epsilon_r\epsilon_s}
-\Gamma_{psn}\Gamma^n_{\;\;\;qr}
(-1)^{\epsilon_s\epsilon_q}
\\ \nonumber
=& -\Gamma_{pqs,r}+
\Gamma_{pqn}\Gamma^n_{\;\;\;sr}+
\Gamma_{pns}\Gamma^n_{\;\;\;qr}
(-1)^{\epsilon_s(\epsilon_n+\epsilon_q)}
+\Gamma_{nqs}\Gamma^n_{\;\;\;pr}
(-1)^{(\epsilon_s+\epsilon_q)(\epsilon_n+\epsilon_p)}
\\ \nonumber &
+\Gamma_{prn}\Gamma^n_{\;\;\;qs}
(-1)^{\epsilon_r(\epsilon_q+\epsilon_s)}
+\Gamma_{pqr,s}
(-1)^{\epsilon_r\epsilon_s}-
\Gamma_{pqn}(\Gamma^n_{\;\;\;rs}
(-1)^{\epsilon_r\epsilon_s}
\\ \nonumber &
-\Gamma_{pnr}\Gamma^n_{\;\;\;qs}
(-1)^{\epsilon_r(\epsilon_n+\epsilon_q)+\epsilon_r\epsilon_s}
-\Gamma_{nqr}\Gamma^n_{\;\;\;ps}
(-1)^{(\epsilon_r+\epsilon_q)(\epsilon_n+\epsilon_p)+\epsilon_r\epsilon_s}
-\Gamma_{psn}\Gamma^n_{\;\;\;qr}
(-1)^{\epsilon_s\epsilon_q}\\
\nonumber
=&
-\Gamma_{pqs,r}+
\Gamma_{nqs}\Gamma^n_{\;\;\;pr}
(-1)^{(\epsilon_s+\epsilon_q)(\epsilon_n+\epsilon_p)}
+\Gamma_{pqr,s}
(-1)^{\epsilon_r\epsilon_s}-
\Gamma_{nqr}\Gamma^n_{\;\;\;ps}
(-1)^{(\epsilon_r+\epsilon_q)(\epsilon_n+\epsilon_p)+\epsilon_r\epsilon_s}
\\
\nonumber
=&
-\Gamma_{pqs,r}+\Gamma_{pqr,s}(-1)^{\epsilon_r\epsilon_s}
-\Gamma_{npr}\Gamma^n_{\;\;\;qs}
(-1)^{\epsilon_r(\epsilon_n+\epsilon_q+\epsilon_s)+\epsilon_n\epsilon_p}+
\Gamma_{nps}\Gamma^n_{\;\;\;qr}
(-1)^{\epsilon_s(\epsilon_n+\epsilon_q)+\epsilon_n\epsilon_p}\\
\nonumber
=&\;R_{pqsr}.
\end{align}

The relations above are derived at $p$. However, since $p\in M$ is arbitrary
the following relations hold in any local coordinate system $(x)$:
the curvature tensor reads in terms of covariant derivatives of Christoffel symbols,
\begin{eqnarray}
\label{Rcd}
R_{ijkl}&=&
-\Gamma_{ijk;l}+
\Gamma_{iln}\Gamma^n_{\;\;\;jk}
(-1)^{\epsilon_l(\epsilon_j+\epsilon_k)}
+\Gamma_{ijl;k}
(-1)^{\epsilon_k\epsilon_l}
-\Gamma_{ikn}\Gamma^n_{\;\;\;jl}
(-1)^{\epsilon_j\epsilon_k},
\end{eqnarray}
and from (\ref{Chtr5}) and (\ref{Rtr}) it follows that the relation (\ref{RA1}) is to be generalized as
\begin{eqnarray}
\label{RAx}
\Gamma_{ijk;l}-\frac{1}{3}Z_{ijkl}=
-\frac{1}{3}[R_{ijkl}+R_{ikjl}(-1)^{\epsilon_j\epsilon_k}].
\end{eqnarray}
The last equation gets an identity when using the definition (\ref{Zx}) of
$Z_{ijkl}(x)$ and relation (\ref{Rcd}) for $R_{ijkl}(x)$ on the r.h.s..

\section{Second order affine extension of symplectic structure
and curvature tensor on Fedosov supermanifolds}

Now, let us consider the relation between the second order affine extension of symplectic structure and the symplectic curvature tensor. It is easily found by taking into account (\ref{wA1}) and (\ref{RA1}). Indeed we obtain
\begin{align}
\label{wAR1}
\nonumber
\Omega_{ij,kl}
&=
A_{ijkl}- A_{jikl}
(-1)^{\epsilon_i\epsilon_j}\\
\nonumber
&=-\frac{1}{3}
\left[\left(R_{ijkl}(0)+R_{ikjl}(0)(-1)^{\epsilon_k\epsilon_j}\right)-
\left(R_{jikl}(0)(-1)^{\epsilon_i\epsilon_j}+
R_{jkil}(0)(-1)^{\epsilon_i(\epsilon_j+\epsilon_k)}\right)\right]\\
\nonumber
&=-\frac{1}{3}
\left[R_{ikjl}(0)(-1)^{\epsilon_k\epsilon_j}-
R_{jkil}(0)(-1)^{\epsilon_i(\epsilon_j+\epsilon_k)}\right]\\
\nonumber
&=-\frac{1}{3}
\left[R_{kilj}(0)(-1)^{\epsilon_i\epsilon_j}
+R_{kjil}(0)(-1)^{\epsilon_l\epsilon_j}\right]
(-1)^{(\epsilon_i+\epsilon_j)\epsilon_k+(\epsilon_i+\epsilon_l)\epsilon_j}\\
&=\frac{1}{3} R_{klji}(0)(-1)^{\epsilon_l\epsilon_i}
(-1)^{(\epsilon_i+\epsilon_j)\epsilon_k+(\epsilon_i+\epsilon_l)\epsilon_j},
\end{align}
where in the last line the Jacobi identity (\ref{Rjac1}) has been used
for the curvature tensor.  Then from (\ref{wAR1}) it follows
\begin{eqnarray}
\label{wAR}
\omega_{ij,kl}(0)=
\frac{1}{3}R_{klij}(0)
(-1)^{(\epsilon_i+\epsilon_j)(\epsilon_k+\epsilon_l)}.
\end{eqnarray}

Obviously, Eq.~(\ref{wAR}) holds at $p\in M$ in normal coordinates $(y)$.
However, it can easily be generalized to arbitrary local coordinates $(x)$.
One only has to remind that Eq.~(\ref{RAx}) is the generalization of Eq.~(\ref{RA1}) for $\Gamma_{ijk,l}(0)$ in order to obtain
\begin{eqnarray}
\label{wxt}
\nonumber
\omega_{ij,kl}(0)
&=&
\Gamma_{ijk,l}(0)-\Gamma_{jik,l}(0)
(-1)^{\epsilon_i\epsilon_j}\\
&=&\omega_{ij,k;l}(x_0)-\frac{1}{3}Z_{ijkl}(x_0)+
\frac{1}{3}Z_{ijlk}(x_0)(-1)^{\epsilon_i\epsilon_j},
\end{eqnarray}
with
\begin{eqnarray}
\label{wxt0}
\omega_{ij,k;l}= \Gamma_{ijk;l}-\Gamma_{jik;l}
(-1)^{\epsilon_i\epsilon_j},
\end{eqnarray}
as a consequence of Eq.~(\ref{sc}).
Thereby, the first derivative of $\omega_{ij}$ is understood as usual
partial one while the second one is the covariant derivative
$\omega_{ij,k;l}\equiv (\omega_{ij,k})\nabla_l$.
Again, since $p\in M$ is arbitrary, we finally obtain the desired
generalization of Eq.~(\ref{wAR}):
\begin{eqnarray}
\label{wARx}
\omega_{ij,k;l}-\frac{1}{3}\left[Z_{ijkl}-Z_{jikl}
(-1)^{\epsilon_i\epsilon_j}\right]=
\frac{1}{3}(-1)^{(\epsilon_i+\epsilon_j)(\epsilon_k+\epsilon_l)}
R_{klij}.
\end{eqnarray}

\section{Third order affine extension of symplectic structure
and first order affine extension of curvature tensor}

First, note that, because of $\omega_{ij;k} =0$ and $\omega_{ln}R^n_{\;\;mjk;i}=(\omega_{ln}R^n_{\;\;mjk})_{;i}=R_{lmjk;i}$,
on any Fedosov supermanifolds the second Bianchi identity exists \cite{gl2},
\begin{eqnarray}
\label{2BI}
R_{ijkl;m}(-1)^{\epsilon_m\epsilon_k}+
R_{ijlm;k}(-1)^{\epsilon_l\epsilon_k}+
R_{ijmk;l}(-1)^{\epsilon_m\epsilon_l}\equiv 0.
\end{eqnarray}

However, there exist further relations among the affine extensions of
$\Gamma_{ijk}$, $R_{ijkl}$ and $\omega_{ij}$ on a Fedosov supermanifold.
In order to show them, let us rewrite $R_{ijkl}$ in terms of
$\Gamma_{ijk}$ and their first derivatives:
\begin{align}
\label{Rn}
\nonumber
R_{ijkl}=&-\Gamma_{ijk,l}+\Gamma_{ijl,k}(-1)^{\epsilon_k\epsilon_l}\\
\nonumber
&+\left(\Gamma_{ikr}\omega^{rs}\Gamma_{sjl}(-1)^{\epsilon_j\epsilon_k}
-\Gamma_{ilr}\omega^{rs}\Gamma_{sjk}
(-1)^{\epsilon_l(\epsilon_j+\epsilon_k)}\right)
(-1)^{\epsilon_r+\epsilon(\omega)(\epsilon_r+\epsilon_s)}\\
&-\left(\omega_{ir}\omega^{rs}_{\;\;\;\;,l}\Gamma_{sjk}
(-1)^{\epsilon_l(\epsilon_j+\epsilon_k+\epsilon_s)}-
\omega_{ir}\omega^{rs}_{\;\;\;\;,j}\Gamma_{sjl}
(-1)^{\epsilon_k(\epsilon_j+\epsilon_s)}\right)
(-1)^{\epsilon_s+
\epsilon(\omega)(\epsilon_r+\epsilon_s)}.
\end{align}
Differentiating both sides w.r.t.~$y^m$, taking
the limit $y\rightarrow 0$ and observing that, because of $\Gamma_{ijk}(0)=0$,
the first extension of the symplectic structure vanishes,$\omega_{ij,k}(0)=0$, we have
\begin{eqnarray}
\label{RA2}
R_{ijkl,m}(0)=-A_{ijklm} + A_{ijlkm}(-1)^{\epsilon_l\epsilon_k}.
\end{eqnarray}
That relation will be used to eliminate within the relation
(\ref{r6}) all the extensions of the Christoffel symbols in favor of
$A_{ijklm}$ and corresponding derivations of symplectic curvature tensor
according to the following relations:
\begin{eqnarray}
\nonumber
A_{ijlkm}(-1)^{\epsilon_l\epsilon_k}
&=&
R_{ijkl,m}(0)+A_{ijklm},
\\ \nonumber
A_{ijmkl}(-1)^{\epsilon_m\epsilon_k}
&=&
R_{ijkm,l}(0)+A_{ijkml}=R_{ijkm,l}(0)+A_{ijklm}(-1)^{\epsilon_l\epsilon_m},
\\ \nonumber
A_{ikljm}(-1)^{\epsilon_j\epsilon_l}
&=&
R_{ikjl,m}(0)+A_{ikjlm}=R_{ikjl,m}(0)+A_{ijklm}(-1)^{\epsilon_j\epsilon_k},
\\ \nonumber
A_{ikmjl}(-1)^{\epsilon_j\epsilon_m}
&=&
R_{ikjm,l}(0)+A_{ikjml}=
R_{ikjm,l}(0) + A_{ijklm}(-1)^{\epsilon_j\epsilon_k+\epsilon_m\epsilon_l},
\\ \nonumber
A_{ilmjk}(-1)^{\epsilon_j\epsilon_m}
&=&
R_{iljm,k}(0)+A_{iljmk}
= R_{iljm,k}(0) + A_{ijlkm}(-1)^{\epsilon_j\epsilon_l+\epsilon_m\epsilon_k}
\\ \nonumber
&=&R_{iljm,k}(0)+R_{ijkl,m}(0)
(-1)^{\epsilon_j\epsilon_l+\epsilon_m\epsilon_k+\epsilon_l\epsilon_k}+
A_{ijklm}
(-1)^{\epsilon_j\epsilon_l+\epsilon_m\epsilon_k+\epsilon_l\epsilon_k}.
\end{eqnarray}
Putting them into the identity (\ref{r6}) we obtain
\begin{align}
\label{r65}
\nonumber
6A_{ijklm}&+2R_{ijkl,m}(0) +
R_{ijkm,l}(0)(-1)^{\epsilon_m\epsilon_l}+
R_{ikjl,m}(0)(-1)^{\epsilon_j\epsilon_k}\\
&+R_{ikjm,l}(0)(-1)^{\epsilon_j\epsilon_k+\epsilon_m\epsilon_l}+
R_{iljm,k}(0)(-1)^{\epsilon_j\epsilon_l+\epsilon_k(\epsilon_m+\epsilon_l)}=0,
\end{align}
and we obtain the
following representation of the second order affine
extension of $\Gamma_{ijk}$ in terms of first order derivatives of
the curvature tensor
\begin{eqnarray}
\label{A2R}
\nonumber
A_{ijklm}
&=&
-\frac{1}{6}\Big[\,
2R_{ijkl,m}(0) +
R_{ijkm,l}(0)(-1)^{\epsilon_m\epsilon_l}+
R_{ikjl,m}(0)(-1)^{\epsilon_j\epsilon_k}\\
&& \qquad + R_{ikjm,l}(0)(-1)^{\epsilon_j\epsilon_k+\epsilon_m\epsilon_l}+
R_{iljm,k}(0)(-1)^{\epsilon_j\epsilon_l+\epsilon_k(\epsilon_m+\epsilon_l)}
\Big].
\end{eqnarray}

In its turn from (\ref{wA}) we have
\begin{eqnarray}
\label{wA3}
\omega_{ik,jlm}(0)=A_{ikjlm}-A_{kijlm}(-1)^{\epsilon_i\epsilon_k},
\end{eqnarray}
and therefore we get
\begin{align}
\label{wA3R}
&\omega_{ij,klm}(0)=-\frac{1}{6}\Big[
 R_{ikjl,m}(0)(-1)^{\epsilon_j\epsilon_k}
+R_{ikjm,l}(0)(-1)^{\epsilon_j\epsilon_k+\epsilon_m\epsilon_l}
+R_{iljm,k}(0)(-1)^{\epsilon_j\epsilon_l+\epsilon_k(\epsilon_l+\epsilon_m)}
\\ \nonumber 
& \qquad\quad
-R_{jkim,l}(0)(-1)^{\epsilon_i(\epsilon_k+\epsilon_j)}
-R_{jkim,l}(0)(-1)^{\epsilon_m\epsilon_l+\epsilon_i(\epsilon_j+\epsilon_k)}
-R_{jlim,k}(0)(-1)^{\epsilon_k(\epsilon_m+\epsilon_l)
+\epsilon_i(\epsilon_j+\epsilon_l)}\Big]
\end{align}
as the representation of the third order affine extensions of
$\omega_{ij}$ in terms of the first order affine extension of the symplectic curvature tensor.

Now, we consider the consequences which follow from the symmetry properties
of $\omega_{ik,jlm}$,
\begin{eqnarray}
\label{ws3}
\omega_{ik,jlm}=\omega_{ik,ljm}(-1)^{\epsilon_l\epsilon_j}
\end{eqnarray}
and the possibility to express $A_{ijklm}$ in terms of the
first order derivative of the curvature tensor (\ref{A2R}).
We have
\begin{align}
\label{ws31}
\nonumber
0=& \,\omega_{ik,jlm}-\omega_{ik,ljm}(-1)^{\epsilon_l\epsilon_j}\\
\nonumber
=&\,A_{ikjlm}-A_{kijlm}(-1)^{\epsilon_i\epsilon_k} -
(-1)^{\epsilon_l\epsilon_j}(A_{ikljm}-A_{kiljm}
(-1)^{\epsilon_i\epsilon_k})\\
=&\,A_{ijklm}(-1)^{\epsilon_j\epsilon_k}-
A_{kjilm}(-1)^{\epsilon_i(\epsilon_k+\epsilon_j)}
-A_{ilkjm}(-1)^{\epsilon_l(\epsilon_k+\epsilon_j)}+
A_{klijm}(-1)^{\epsilon_i\epsilon_k+\epsilon_l(\epsilon_i+\epsilon_j)}.
\end{align}
Using Eqs.~(\ref{wA3}) and the symmetry properties of the curvature tensor
we can be rewritten as
\begin{align}
\label{ws32}
\nonumber
0=&\;
2R_{ijkl,m}(0)(-1)^{\epsilon_j\epsilon_k}-
2R_{kjil,m}(0)(-1)^{\epsilon_i(\epsilon_j+\epsilon_k)}-
2R_{ilkj,m}(0)(-1)^{\epsilon_l(\epsilon_j+\epsilon_k)}\\
\nonumber
&+
2R_{klij,m}(0)(-1)^{\epsilon_i(\epsilon_l+\epsilon_k)+\epsilon_l\epsilon_j}
+R_{ijkm,l}(0)(-1)^{\epsilon_k\epsilon_j+\epsilon_m\epsilon_l}
-R_{kjim,l}(0)(-1)^{\epsilon_i(\epsilon_k+\epsilon_j)+\epsilon_m\epsilon_l}\\
\nonumber
&-R_{ilkm,j}(0)(-1)^{\epsilon_j(\epsilon_m+\epsilon_l)+\epsilon_k\epsilon_l}
 +R_{klim,j}(0)(-1)^{\epsilon_j(\epsilon_m+\epsilon_l)+
  \epsilon_i(\epsilon_k+\epsilon_l)}\\
\nonumber
&+R_{iljm,k}(0)(-1)^{\epsilon_j(\epsilon_k+\epsilon_l)+
  \epsilon_k(\epsilon_m+\epsilon_l)}
 -R_{kljm,i}(0)(-1)^{\epsilon_j(\epsilon_i+\epsilon_l)+
  \epsilon_i(\epsilon_k+\epsilon_l+\epsilon_m}\\
&-R_{ijlm,k}(0)(-1)^{\epsilon_k(\epsilon_j+\epsilon_m+\epsilon_k)}
 +R_{kjlm,i}(0)(-1)^{\epsilon_i(\epsilon_j+\epsilon_m+\epsilon_k+\epsilon_l)}.
\end{align}
Due to the first Bianchi identities (\ref{1BI}) for the curvature tensor
the first four terms in (\ref{ws32}) vanish:
\begin{align}
\nonumber
R_{ijkl,m}&(-1)^{\epsilon_j\epsilon_k}
-R_{kjil,m}(-1)^{\epsilon_i(\epsilon_j+\epsilon_k)}
-R_{ilkj,m}(-1)^{\epsilon_l(\epsilon_j+\epsilon_k)}
+R_{klij,m}(-1)^{\epsilon_i(\epsilon_l+\epsilon_k)+\epsilon_l\epsilon_j}
\\ \nonumber &
=(-1)^{\epsilon_i\epsilon_l+\epsilon_j\epsilon_k)}
\big[
 R_{ijkl}(-1)^{\epsilon_i\epsilon_l}
+R_{jkli}(-1)^{\epsilon_i(\epsilon_j+\epsilon_k)}
\\ \nonumber & \qquad\qquad\qquad
+R_{lijk}(-1)^{\epsilon_l(\epsilon_j+\epsilon_k)}
+R_{klij}(-1)^{\epsilon_j(\epsilon_k+\epsilon_l)+\epsilon_i\epsilon_k)}
\big]_{,m}
=0
\end{align}

The remaining terms in (\ref{ws32}) can be presented in the form
\begin{align}
\label{ws33}
\nonumber
&\big[
R_{jikm,l}(0)(-1)^{\epsilon_m\epsilon_i}+
R_{jkim,l}(0)(-1)^{\epsilon_i\epsilon_k}\big]
(-1)^{\epsilon_j(\epsilon_i+\epsilon_k)+\epsilon_m(\epsilon_i+\epsilon_l)}
\\
\nonumber
+&\big[
R_{limk,j}(0)(-1)^{\epsilon_i\epsilon_k}+
R_{lkim,j}(0)(-1)^{\epsilon_k\epsilon_m}\big]
(-1)^{\epsilon_i(\epsilon_k+\epsilon_l)+
\epsilon_j(\epsilon_m+\epsilon_l)+\epsilon_k(\epsilon_i+\epsilon_l)}\\
\nonumber
+&\big[
R_{iljm,k}(0)(-1)^{\epsilon_m\epsilon_l}+
R_{ijml,k}(0)(-1)^{\epsilon_j\epsilon_l}\big]
(-1)^{\epsilon_l(\epsilon_k+\epsilon_j)+
\epsilon_k(\epsilon_m+\epsilon_j)+\epsilon_m\epsilon_l)}\\
+&\big[
R_{klmj,i}(0)(-1)^{\epsilon_l\epsilon_j}+
R_{kjlm,i}(0)(-1)^{\epsilon_j\epsilon_m}\big]
(-1)^{\epsilon_i(\epsilon_l+\epsilon_m)+
\epsilon_i(\epsilon_k+\epsilon_j)+\epsilon_m\epsilon_j}=0.
\end{align}
Taking into account the Jacobi identity (\ref{Rjac1})
this equation can be rewritten as
\begin{align}
\label{ws34}
\nonumber
0=&\;R_{jmik,l}(0)(-1)^{\epsilon_m\epsilon_k
+\epsilon_j(\epsilon_i+\epsilon_k)+\epsilon_m(\epsilon_i+\epsilon_l)}
+R_{lmki,j}(0)(-1)^{\epsilon_i\epsilon_m+\epsilon_l(\epsilon_i+\epsilon_k)+
\epsilon_j(\epsilon_m+\epsilon_l)}
\\
&\!\!+R_{imlj,k}(0)(-1)^{\epsilon_m\epsilon_j
+\epsilon_l(\epsilon_k+\epsilon_j+\epsilon_m)+
\epsilon_k(\epsilon_m+\epsilon_j)}
+R_{kmjl,i}(0)(-1)^{\epsilon_l\epsilon_m
+\epsilon_i(\epsilon_l+\epsilon_m)+
\epsilon_i(\epsilon_k+\epsilon_j)+\epsilon_m\epsilon_j}.
\end{align}
Finally we obtain the following identities for the first affine
extension of the curvature tensor
\begin{align}
\label{IIRy}
\nonumber
&R_{mjik,l}(0)(-1)^{\epsilon_j(\epsilon_i+\epsilon_k)}-
R_{mijl,k}(0)(-1)^{\epsilon_k(\epsilon_l+\epsilon_j)}\\
&+R_{mkjl,i}(0)(-1)^{\epsilon_i(\epsilon_j+\epsilon_k+\epsilon_l)}-
R_{mlik,j}(0)(-1)^{\epsilon_l(\epsilon_i+\epsilon_j+\epsilon_k)}=0.
\end{align}
In local coordinates $(x)$ the partial derivatives have to be replaced by the covariant ones and these identities evidently have the form
\begin{align}
\label{IIRx}
&R_{mjik;l}(-1)^{\epsilon_j(\epsilon_i+\epsilon_k)}
-R_{mijl;k}(-1)^{\epsilon_k(\epsilon_l+\epsilon_j)}
+R_{mkjl;i}(-1)^{\epsilon_i(\epsilon_j+\epsilon_k+\epsilon_l)}
-R_{mlik;j}(-1)^{\epsilon_l(\epsilon_i+\epsilon_j+\epsilon_k)}=0.
\end{align}

\section{Discussion}
In this paper we continued investigation of properties of even and odd
Fedosov supermanifolds which we began in \cite{gl2}.

Using normal coordinates on a supermanifold equipped with a
symmetric connection we have derived infinite system of equations
for affine extension of the Christoffel symbols (\ref{rn}). With
the help of these equations we have found relations among the
first order affine extensions of the Christoffel symbols and the
curvature tensor (\ref{RA1}), the second order affine extension of
symplectic structure and the curvature tensor (\ref{wAR}), the
third order affine extension of symplectic structure and the first
order affine extension of the curvature tensor (\ref{wA3R})
existing in normal coordinates on any Fedosov supermanifold. In
similar way  it is possible to find relations containing higher
order affine extensions of sypmlectic structure, the Christoffel
symbols and the curvature tensor. This procedure is closely
connected with relations (\ref{rn}), (\ref{wA}), (\ref{Rn}). We
have established the form of the obtained relations in any local
coordinates (see (\ref{RAx}), (\ref{wARx})). It was shown  the
tensor field $\Gamma_{ijk;l}(x)-1/3Z_{ijkl}(x)$ in terms of what
the relations obtained can be presented (\ref{Chtr5}). We have
derived a specific identity (\ref{IIRx}) for the first covariant
derivative of the curvature tensor (\ref{R}).

\vspace{0.5cm}

\noindent
{\Large{\bf {Acknowledgments}}}

\vspace{0.5cm} \noindent PL would like to thank Dieter L\"ust for
his very warm hospitality at the Institute for Physics, the
Humboldt University of Berlin, where this work has been started.
We would like also to thank S. Waldmann for his remarks concerning
deformation quantization on arbitrary even symplectic
supermanifolds. The work was supported by Deutsche
Forschungsgemeinschaft (DFG) grants DFG 436 RUS 17/15/04. The work
of PL was also supported under grant DFG 436 RUS 113/669/0-2,
Russian Foundation for Basic Research (RFBR) grants 03-02-16193
and 04-02-04002, the President grant LSS 1252.2003.2 and INTAS
grant 03-51-6346.

\end{document}